\documentclass[twoside,twocolumn,english,aps,superscriptaddress,prb]{revtex4}
\pdfoutput=1
\usepackage[latin9]{inputenc}
\usepackage{amsmath}
\usepackage{amssymb}
\usepackage{graphicx}
\usepackage{color}

\newcommand{\unit}{1\!\!1}

\begin{document}

\title{Anderson localization and momentum-space entanglement}


\author{Eric C. Andrade}
\affiliation{Institut für Theoretische Physik, Technische Universität Dresden,
01062 Dresden, Germany}
\affiliation{Instituto de Física Teórica, Universidade Estadual Paulista, Rua
Dr. Bento Teobaldo Ferraz, 271 - Bl. II, 01140-070, São Paulo, SP,
Brazil}

\author{Mark Steudtner}
\affiliation{Institut für Theoretische Physik, Technische Universität Dresden,
01062 Dresden, Germany}

\author{Matthias Vojta}
\affiliation{Institut für Theoretische Physik, Technische Universität Dresden,
01062 Dresden, Germany}


\begin{abstract}
We consider Anderson localization and the associated metal--insulator transition for non-interacting fermions in $D=1,2$ space dimensions in the presence of spatially correlated on-site random potentials. To assess the nature of the wavefunction, we follow a recent proposal to study momentum-space entanglement. For a $D=1$ model with long-range disorder correlations, both the entanglement spectrum and the entanglement entropy allow us to clearly distinguish between extended and localized states based upon a single realization of disorder. However, for other models including the $D=2$ case with long-range correlated disorder, we find that the method is not similarly successful. We analyze the reasons for its failure, concluding that the much desired generalization to higher dimensions may be problematic.
\end{abstract}

\date{\today}

\maketitle


\section{Introduction}

The possibility that electronic bound states can be formed in presence
of a random potential was first discussed in Anderson's pioneering
work,\cite{anderson58} where it was shown that, for a sufficiently
strong randomness, single-particle wavefunctions can become exponentially
localized, leading to a sharp metal-insulator transition\cite{book-vlad}
at $T=0$.

For non-interacting fermions with uncorrelated random potentials,
and in absence of spin-orbit coupling, the scaling theory of localization\cite{abrahams79}
predicts that all single-particle states are localized for dimensions
$D\le2$ for any amount of disorder, and thus the Anderson metal-insulator
transition only takes place in $D>2$. One way to circumvent the scaling
predictions and to observe an Anderson transition in low-dimensional
systems is to consider systems with spatially correlated random potentials.\cite{dunlap90,moura98,izrailev99,garcia-garcia09,santos07,sil08,croy11}
Such correlated disorder potential could emerge, for example, as an
effective description of interacting electronic system in the presence
of impurities, where the Fermi liquid readjusts itself, producing
a spatially inhomogeneous pseudopotential ``seen'' by quasiparticles.\cite{herbut01,andrade10}

Over the years, several quantities which probe the extension of the
electronic wave function were employed to theoretically study the
Anderson metal-insulator transition. These include the local density
of states,\cite{tmt} the participation ratio,\cite{wegner80,johri12}
the Lyapunov exponent,\cite{farchioni92} the localization length,\cite{mackinnon81}
and the conductance,\cite{anderson80} among others.\cite{markos06}
More recently, the concept of quantum entanglement has been successfully
applied to the study of the Anderson transition.\cite{jia08,chen12}
This success is mainly due to the fact that entanglement and the spatial
extension of the electronic wavefunctions are closely related: Extended
states are entangled in position space, whereas localized states are
not.

While the use of entanglement in position space appears natural, the
authors of Ref.~\onlinecite{Mondragon-Shem13} recently showed
that the entanglement in {\em momentum}-space\cite{thomale10,lundgren12} also appears to be
a useful tool to study the presence of extended states. As an example,
the authors revisited the so-called random dimer model\cite{dunlap90}
in $D=1$ and pointed out that a direct analysis of the entanglement
spectrum, constructed from a momentum-space partition, provides a
sharper distinction between the extended and localized states as compared
to the position-space partition. Remarkably, useful and accurate results
could be obtained from a single realization of disorder. If generalizable to
$D>1$, this would make the method particularly useful considering
that the numerical effort increases drastically for higher $D$ on
top of which conventional methods to study localization require extensive
averaging over different realizations of disorder.

Motivated by the success of Ref.~\onlinecite{Mondragon-Shem13},
we ask whether momentum-space entanglement as a probe of localization
is of wider applicability. To this end, we apply this concept to other
low-dimensional non-interacting models with correlated disorder. First,
we show that for a $D=1$ model with long-range correlated disorder
momentum-space entanglement indeed provides a clear characterization
of the Anderson transition. Second, we study a model of coupled chains
which displays coexisting extended and localized states, and we also
investigate the Anderson transition for correlated disorder in $D=2$.
In both cases we find that the proposed method fails to capture the
presence of extended states, casting doubts on the general utility
of momentum-space entanglement in the context of localization phenomena.
We suggest two reasons for the failure, one related to the specific
construction of entanglement measures employed in Ref.~\onlinecite{Mondragon-Shem13}
and another one related simply to the topology of higher-dimensional
momentum space.

The body of the paper is organized as follows: In Sec. \ref{sec:model-entang}
we briefly present the tight-binding model studied by Anderson\cite{anderson58}
and discuss the calculation of entanglement spectrum and entropy.
In Sec. \ref{sec:1d} we investigate the metal-insulator transition
in a $D=1$ model with long-range correlated disorder. In Sec. \ref{sec:ladder}
we study a particular model of coupled chains with correlated disorder
within a unit cell. We then return to long-range correlated disorder,
in Sec. \ref{sec:2d}, but now considering $D=2$. A discussion of
methodological aspects and possible reasons for the apparent failure
of momentum-space entanglement is in Sec.~\ref{sec:discussion}.
A short summary concludes the paper.


\section{\label{sec:model-entang}Disordered Hamiltonian and entanglement}

To study the Anderson transition, we consider a single-band tight-binding
model of spinless fermions on a lattice with $N$ sites
\begin{eqnarray}
\mathcal{H} & = & \sum_{i}\varepsilon_{i}n_{i}-t\sum_{\left\langle ij\right\rangle }\left(c_{i}^{\dagger}c_{j}+c_{j}^{\dagger}c_{i}\right),\label{eq:tbh}
\end{eqnarray}
 where $t$ is the hopping matrix element between nearest-neighbor
sites $\langle ij\rangle$, $c_{i}^{\dagger}$$\left(c_{i}\right)$
is the creation (annihilation) operator of an electron at site $i$,
$n_{i}=c_{i}^{\dagger}c_{i}$ is the number operator, and $\varepsilon_{i}$
are the site energies forming the disorder potential. We measure all
our energies in units of $t$ and consider periodic boundary conditions.

\subsection{Entanglement via correlation matrix}

To investigate Anderson localization, we use the notion of quantum
entanglement and follow Ref.~\onlinecite{Mondragon-Shem13}. We
begin by partitioning the system into two regions $\mathcal{A}$ and
$\mathcal{B}$ which may correspond either to a subset of sites (real-space
partitions) or a subset of single-particle momenta (momentum-space
partitions). The next step is to construct a reduced density operator
$\rho_{\mathcal{A}}$ which only acts in the many-body Hilbert space
defined by $\mathcal{A}$ and, by construction, reproduces all expectation
values in region $\mathcal{A}$. From $\rho_{\mathcal{A}}$ we obtain
the entanglement spectrum and entropy, which quantify how much information
region $\mathcal{A}$ contains about the physics in region $\mathcal{B}$.

Any reduced density operator can be written in terms of an entanglement
Hamiltonian $\mathcal{H}_{A}$, $\rho_{A}=\mbox{exp}[-\mathcal{H}_{A}]/Z$,
where $Z$ is a normalization constant. For a model of non-interacting
particles such as Eq.~\eqref{eq:tbh} it can be shown that the entanglement
Hamiltonian $\mathcal{H}_{A}$ has the form a free-fermion Hamiltonian.\cite{peschel03,peschel09}
Its eigenvalues $\varepsilon_{A}^{i}$ are in one-to-one correspondence
to the eigenvalues $\zeta_{i}$ of the two-point correlation matrix
$C_{ij}$ according to\cite{peschel03,peschel09} $\zeta_{i}=\left[\mbox{exp}\left[\varepsilon_{A}^{i}\right]+1\right]^{-1}$
where
\begin{eqnarray}
C_{ij} & = & \langle\Psi|c_{i}^{\dagger}c_{j}|\Psi\rangle,\label{eq:cij}
\end{eqnarray}
 with $\left|\Psi\right\rangle $ being a free-fermion ground state,
i.e., the Fermi sea filled up to the Fermi energy $E_{F}$ and, for
simplicity, we consider $i,j\in\mathcal{A}$. Since the entanglement
Hamiltonian $\mathcal{H}_{A}$ has a free-fermion form, the entanglement
entropy is then simply given by
\begin{eqnarray}
S & = & -\sum_{i}\left(\zeta_{i}\mbox{log}\zeta_{i}+\left(1-\zeta_{i}\right)\mbox{log}\left(1-\zeta_{i}\right)\right),\label{eq:entang-ent}
\end{eqnarray}
 and, from now on, we refer to $\zeta_{i}$ as the entanglement spectrum.
We see that $\zeta_{i}=0,1$ correspond to no entanglement, whereas
$\zeta_{i}=1/2$ corresponds to maximum entanglement. We also point
out that $S$ does not have a fixed upper bound in the present scheme,
since we construct $C_{ij}$ using all states up to the Fermi level.

To numerically evaluate $C_{ij}$, we first write the operator $c_{i}$
in terms of the eigenstates $\left|\nu\right\rangle $ of the single-particle
problem, $c_{i}=\sum_{\nu}\left\langle i\right.\left|\nu\right\rangle a_{\nu}$,
with $\mathcal{H}=\sum_{\nu}\varepsilon_{\nu}a_{\nu}^{\dagger}a_{\nu}$,
and then
\begin{eqnarray}
C_{ij} & = & \sum_{\nu}\left\langle \nu\right.\left|i\right\rangle \left\langle j\right.\left|\nu\right\rangle \theta\left(E_{F}-\varepsilon_{\nu}\right),\label{eq:cij_final}
\end{eqnarray}
 where $\theta\left(x\right)$ is the usual step-function and $E_{F}$
is the Fermi energy.

For a real-space space partition, extended states are expected to
display entanglement whereas localized states are not. A momentum-space
partition was only recently investigated in the context of disordered
electronic system\cite{Mondragon-Shem13} and, conversely to its real-space
counterpart, when looking for extended states we should seek for the
{\em absence} of entanglement since an extended state in real space
has a well defined (localized) momentum.

We note that the partition of the original system into two subregions
will generate spurious (or boundary) entanglement also in situations
where entanglement is expected to be weak: States which are localized
in the vicinity of the cut and have weight both in $\mathcal{A}$
and $\mathcal{B}$ will induce entanglement even though states away
from the cut are not entangled. As we will see later, such spurious
entanglement plays a prominent role in higher dimensions.

\subsection{Correlation matrix vs. single-particle entanglement}

While conventional measures of localization, e.g. the inverse participation
ratio, are defined for single-particle states, the eigenvalues of
the correlation matrix $C_{ij}$ in Eq. \eqref{eq:cij_final}, proposed
as indicator of localization in Ref.~\onlinecite{Mondragon-Shem13},
involve the whole Fermi sea.\cite{pouranvari14} As will become clear in the remainder
of the paper, this has a number of consequences.

First we note that in order to study real-space entanglement there
is no need to consider $C_{ij}$. One may instead focus on single-particle
entanglement\cite{jia08,chen12} which can be directly obtained from
an individual single-particle state.

Second, however, single-particle entanglement can be problematic for
a momentum-space partition. To illustrate this point, we consider
the clean (disorder-free) limit in $D=1$. In this case, the real
single-particle eigenstates of Eq. \eqref{eq:tbh} can be written
as $\psi_{k}^{c}\left(x\right)=\sqrt{2/L}\mbox{\,}\mbox{cos}\left(kx\right)$
or $\psi_{k}^{s}\left(x\right)=\sqrt{2/L}\mbox{\,}\mbox{sin}\left(kx\right)$.
If we now evaluate $C_{ij}$ for a real space partition, the only
non-vanishing contribution to the entanglement spectrum is
$\zeta_{1}=\sum_{i=1}^{L/2}|\psi_{k}^{c\left(s\right)}\left(x\right)|^{2}=1/2$.
This maximally entangled situation simply reflects the fact that the
wavefunctions $\psi_{k}^{c\left(s\right)}\left(x\right)$ are extended
over the entire lattice. Performing a similar calculation for a momentum-space
partition, we obtain $\zeta_{1,2}=1/2$, which wrongly suggests an
ill-defined momentum state and thus a localized wavefunction in real
space. Behind this failure lies the fact that the Fourier-transformed
single-particle states $\psi_{k}^{c\left(s\right)}\left(p\right)$
are non-zero at both $\pm k$. One way to resolve this problem is
to consider two particles, occupying both {$\psi_{k}^{c}$
and $\psi_{k}^{s}$, and to construct $C_{ij}$ for
this two-particle system. In this case, we obtain $\zeta_{1,2}=1$
as the non-vanishing contributions to the momentum-space entanglement spectrum, which
now correctly translates into extended wavefunctions in real space.
Therefore, one role of the Fermi sea in Eq.~\eqref{eq:cij_final}
is to account for the presence of states at $\pm k$ in momentum-space partitions.

Ultimately, the fact that one considers the whole Fermi sea introduces
interference between different occupied states, which in turn can
reveal extra information not present at level of single-particle entanglement.\cite{pouranvari14,prodan10,Mondragon-Shem14}


\section{\label{sec:1d}Long-range correlated disorder in $D=1$}

We first investigate a chain of length $N=L$. To construct a sequence
of correlated site energies, we follow the proposal of Ref.~\onlinecite{moura98}
and consider that the site energies $\varepsilon_{i}$ have a spectral
density $S\left(k\right)\propto1/k^{\alpha}$, where $S\left(k\right)$
is the Fourier transform of the two-point correlation function $\left\langle \varepsilon_{i}\varepsilon_{j}\right\rangle $,
where $\left\langle \cdots\right\rangle $ denotes average over disorder.
When the exponent $\alpha=0$, we recover the usual Anderson\cite{anderson58}
model with uncorrelated disorder, which shows a white noise spectrum
and a local two-point correlation function $\left\langle \varepsilon_{i}\varepsilon_{j}\right\rangle =\left\langle \varepsilon_{i}^{2}\right\rangle \delta_{i,j}$.

To generate a correlated sequence with a given spectral density we
follow a standard procedure\cite{osborne89} and write

\begin{eqnarray}
\varepsilon_{i} & = & \sum_{n_{k}=1}^{L/2}\left[\left(\frac{2\pi}{L}\right)^{1-\alpha}\frac{1}{n_{k}^{\alpha}}\right]\mbox{cos}\left(\frac{2\pi}{L}in_{k}+\varphi_{k}\right),\label{eq:plaw-energies}
\end{eqnarray}
 where $\varphi_{k}$ are $L/2$ independent random variables uniformly
distributed in the interval $\left[0,2\pi\right]$ and $k=2\pi n_{k}/L$.
Due to the randomness of $\varphi_{k}$, we have $\left\langle \varepsilon_{i}\right\rangle =0$
and we normalize the energy such that $\sigma_{\varepsilon}=\sqrt{\left\langle \varepsilon_{i}^{2}\right\rangle }=1$
in order to keep the same disorder strength for all $\alpha$. Interestingly,
the energies generated in this fashion have the property that $\left\langle \varepsilon_{i}\varepsilon_{i+L/2}\right\rangle =2^{1-\alpha}-1$,\cite{petersen13}
meaning that $\varepsilon_{i}$ and $\varepsilon_{i+L/2}$ (the two
most distant sites) are \emph{anti-correlated} for $\alpha>1$. The
resulting site energies are shown in Fig. \ref{fig:plaw-bare-energies}(a)
and we see that the energy spatial profile becomes smoother as $\alpha$
increases and that the anti-correlation between the most distant sites
is more evident.

As first shown in Ref.~\onlinecite{moura98}, this model displays
extended states when the exponent $\alpha>2$, with the location of
the mobility edges $E_{c}$ depending on the value of $\alpha$. Moreover,
because of the long-range correlations in the disorder potential,
this problem lacks self-averaging,\cite{moura98,nishino09} and thus
the location of the phase boundaries in the phase diagram is particular
to the correlated sequence used (see, for instance, Figs. \ref{fig:plaw-bare-energies}(c)
and (d)).

\subsection{Density of states and inverse participation ratio}

We start by investigating the total density of states, identical to
the site-averaged local density of states $\rho_{i}\left(E\right)$,
$\rho_{tot}\left(E\right)=N^{-1}\sum_{i=1}^{N}\rho_{i}\left(E\right)$,
with
\begin{eqnarray}
\rho_{i}\left(E\right) & = & \sum_{\nu=1}^{N}\left|\left\langle i\right.\left|\nu\right\rangle \right|^{2}\delta\left(E-E_{\nu}\right),\label{eq:rho-av}
\end{eqnarray}
 where $\left\langle i\right.\left|\nu\right\rangle $ is the amplitude
of the $\nu$-eigenvector of Eq. \eqref{eq:tbh}, with energy $E_{\nu}$,
at the site located at $r_{i}$. We numerically evaluate the delta
function as $\delta\left(x\right)\approx\Gamma^{-1}\theta\left(\Gamma/2-\left|x\right|\right)$
and we generally choose the width $\Gamma=0.1t$. In Fig. \ref{fig:plaw-bare-energies}(b)
we show $\rho_{tot}$ for different values of $\alpha$. Because of
the imposed normalization $\left\langle \varepsilon_{i}^{2}\right\rangle =1$
the width of $\rho_{tot}$ is essentially independent of $\alpha$.
We also see that $\rho_{tot}\neq0$ for $\left|E\right|\le4t$, with
no sharp band edges. As $\alpha$ increases, the curves become smoother
following the same trend as the site energies. Even though $\rho_{i}$
undergoes a qualitative change upon localization, because it directly
measures the local amplitude of the electronic wave functions, $\rho_{tot}$
shows no sign of an Anderson transition for $\alpha>2$, since site
to site fluctuations are averaged out.\cite{tmt}

We also consider the inverse participation ratio
\begin{eqnarray}
\mbox{IPR}_{\nu} & = & \sum_{i=1}^{N}\left|\left\langle i\right.\left|\nu\right\rangle \right|^{4}.\label{eq:ipr}
\end{eqnarray}
 Generically, we have that $\mbox{IPR}_{\nu}\sim\mathcal{O}\left(1\right)$
for a localized state and $\mbox{IPR}_{\nu}\sim\mathcal{O}\left(1/N\right)$
for an extended one. The value of the $\mbox{IPR}_{\nu}$ averaged
over disorder is often employed as a powerful tool to investigate
the Anderson transition. In the present case, however, due to the
long-range character of the disorder, the position of the mobility
edges changes considerably from sample to sample, Figs. \ref{fig:plaw-bare-energies}(c)
and (d), even in the thermodynamic limit.\cite{nishino09} Due do
this peculiar behavior, we consider the $\mbox{IPR}_{\nu}$ only for
single disorder realizations. Interestingly, for the current model,\cite{nishino09}
the positions of the mobility edges in a given sample are determined
by sharp peaks in the $\mbox{IPR}_{\nu}$, which separate the extended
states ($N\times\mbox{IPR}_{\nu}$ constant) from the localized ones
($N\times\mbox{IPR}_{\nu}$ size dependent), Figs. \ref{fig:plaw-bare-energies}(c)
and (d).

\begin{figure}[t]
\begin{centering}
\includegraphics[scale=0.4]{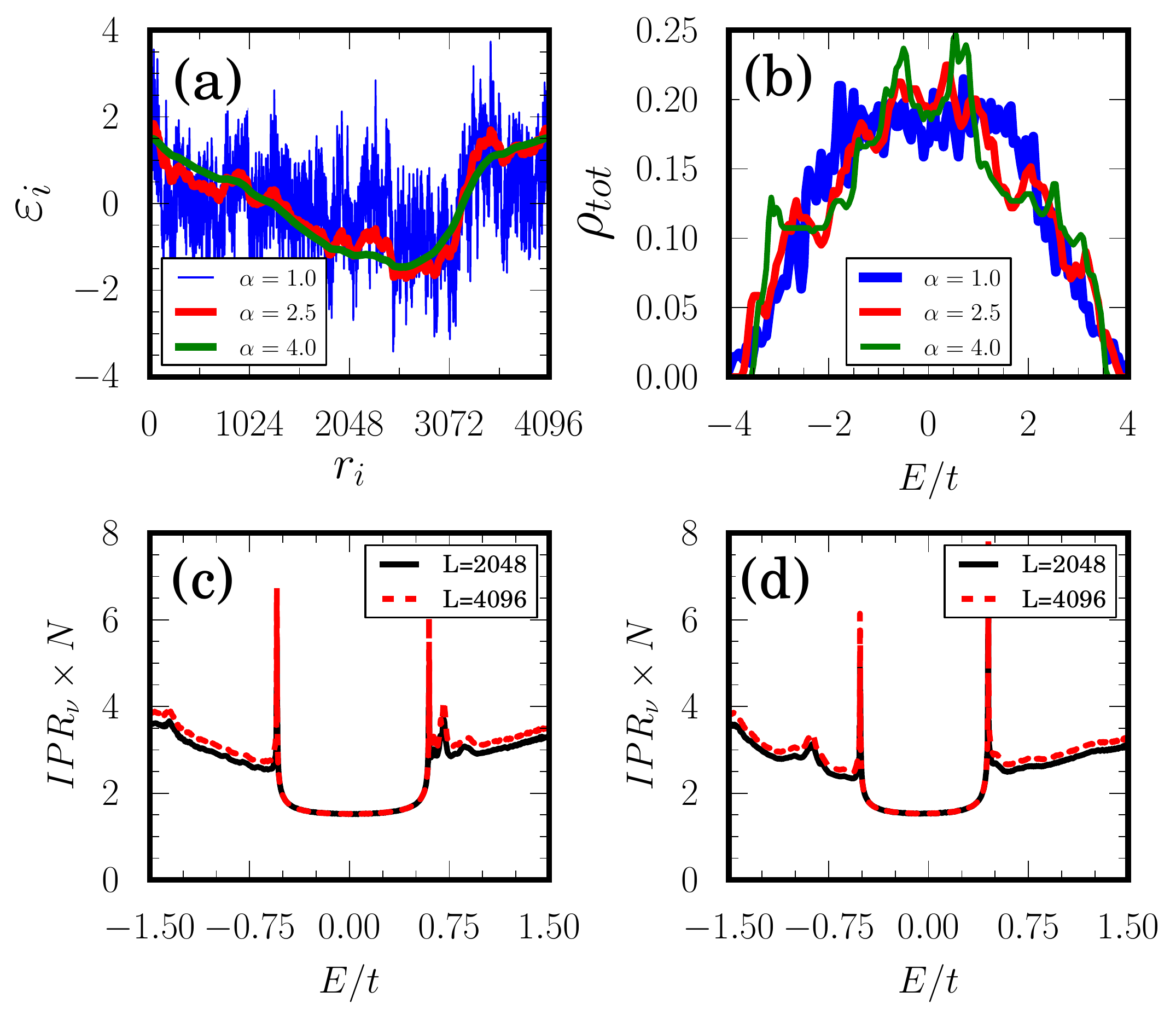}
\par\end{centering}

\caption{\label{fig:plaw-bare-energies} (a) Spatial profile of the correlated
site energies $\varepsilon_{i}$, Eq. \eqref{eq:plaw-energies}, as
a function of the site position $r_{i}$ for three values of $\alpha$.
(b) Total density of states $\rho_{tot}$ for the disordered tight-binding
Hamiltonian Eq. \eqref{eq:tbh} with long-range correlated site disorder
as given by Eq. \eqref{eq:plaw-energies} as a function of the energy
$E$ for three different values of the exponent $\alpha$. Notice
that the fluctuations in both $\rho_{tot}$ and $\varepsilon_{i}$
diminish with $\alpha$ and that $\rho_{tot}$ shows no sign of the
Anderson transition for $\alpha>2$. Here we consider a single realization
of disorder and a chain of size $L=4096$. (c) and (d) Inverse participation
ratio, multiplied by the system size, as a function of energy for
$\alpha=5.0$ and two different system sizes. In (c) and (d) we consider
two distinct realizations of disorder. The sharp peaks in the $\mbox{IPR}_{\nu}$
mark the position of the mobility edges, which are sample dependent.}
\end{figure}

\subsection{Entanglement spectrum and entropy}

We now turn to characterize the Anderson transition via entanglement
measures. To this end, we calculate the entanglement spectrum $\zeta_{i}$
and entropy $S$ for a \emph{single} disorder realization. First,
we consider a partition in real space with $\mathcal{A}\in\left[1,L/2\right]$
and $\mathcal{B}\in\left[L/2+1,L\right]$, with sample results in
Figs. \ref{fig:plaw-spectrum-1d}(a),(d). From previous work,\cite{moura98}
we know that extended states at $E_{F}=0$ occur for $\alpha>2.5$
and, while the entropy saturates to its maximum value in this region,
the entanglement spectrum $\zeta_{i}$ does not display a sharp boundary,
since there is appreciable entanglement also for $\alpha<2.5$.

Second, we consider a cut in momentum space\cite{Mondragon-Shem13}
with $\mathcal{A}\in\left[0,\pi\right]$ and $\mathcal{B}\in\left]\pi,2\pi\right]$
with sample results in Figs. \ref{fig:plaw-spectrum-1d}(b),(c),(d).
For the clean case, this partition corresponds to separation between
the right and left movers. Remarkably, the momentum-space entanglement
spectrum displays a qualitative change as function of $\alpha$. Comparing
Figs. \ref{fig:plaw-spectrum-1d}(a) and \ref{fig:plaw-spectrum-1d}(b)
we see that the momentum-space cut clearly shows the presence of extended
states. We recall that for this cut extended states are associated
with the suppression of entanglement, $\zeta_{i}=0,1$, which in Fig.
\ref{fig:plaw-spectrum-1d}(b) occurs for $\alpha>2.5$. It is also
interesting to notice that the entanglement entropy curve in Fig.
\ref{fig:plaw-spectrum-1d}(d) is very smooth, even though no average
of disorder was performed. For completeness, in Fig. \ref{fig:plaw-spectrum-1d}(c)
we show the momentum-space entanglement spectrum for fixed $\alpha$
as function of the Fermi energy. For $E_{F}$ smaller than the mobility
edge we again see a suppression of entanglement; recall that the position
of the mobility edge is independently known through $\mbox{IPR}_{\nu}$
as in Figs. \ref{fig:plaw-bare-energies}(b) and (d).

Despite of all clear evidences for the presence of extended states
in Figs. \ref{fig:plaw-spectrum-1d}(b),(c),(d), it is difficult to
determine the precise location of the critical $\alpha$, or of the
mobility edges, with the current method, a situation which remains
unchanged even for larger system sizes. As an example, we return to
Figs. \ref{fig:plaw-spectrum-1d}(b),(d) to point out that a critical
value of $\alpha=2.4$ or $2.6$ is as plausible as the quoted value
of $2.5$.\cite{moura98} Thus, when investigating models for which
the precise localization of the transition is unknown, the current
method may have to be complemented by different measures of localization.\cite{markos06}

\begin{figure}[t]
\begin{centering}
\includegraphics[scale=0.4]{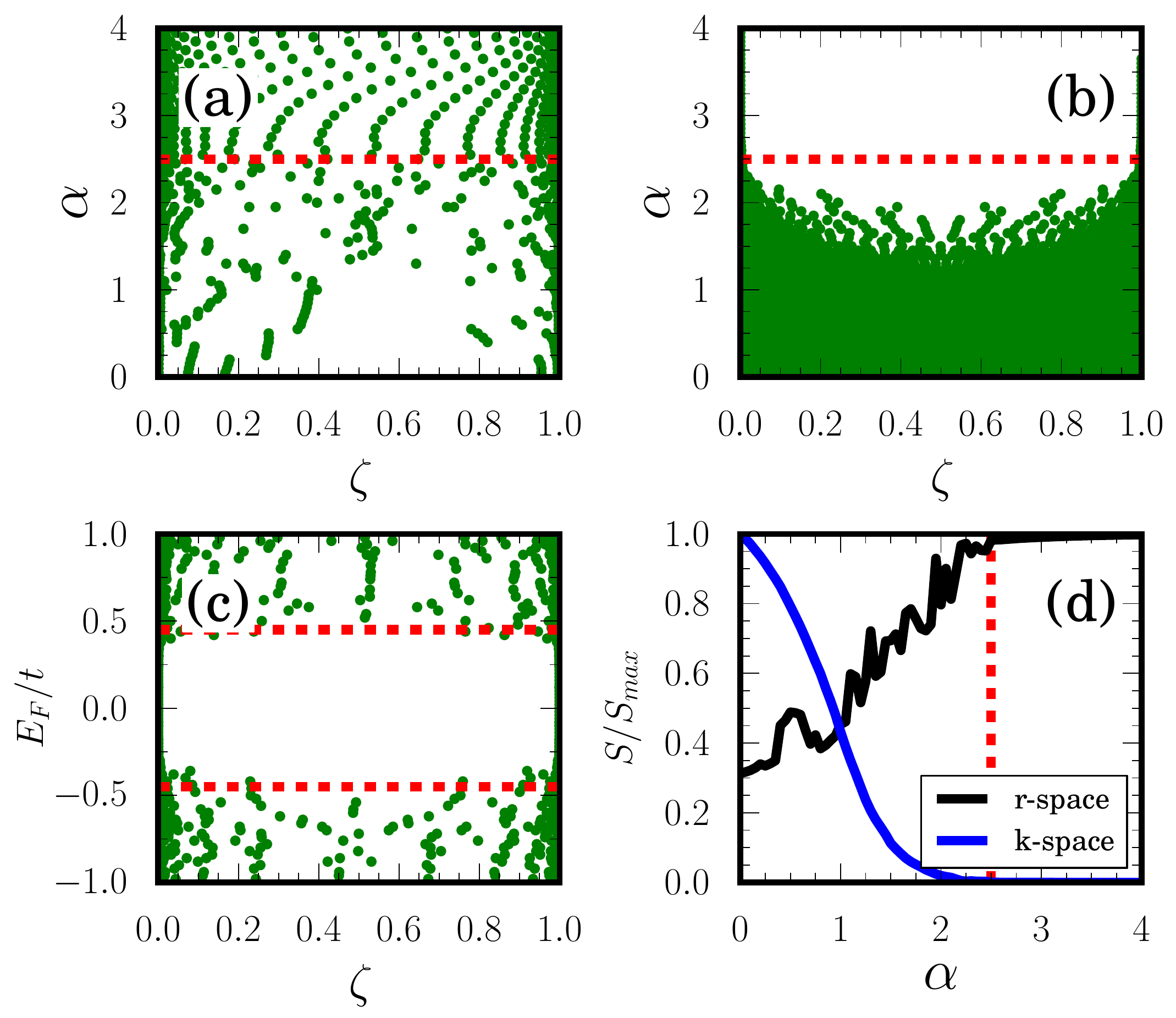}
\par\end{centering}

\caption{\label{fig:plaw-spectrum-1d} Entanglement spectra $\zeta_{i}$ and
entropy $S$ for the disordered tight-binding Hamiltonian Eq. \eqref{eq:tbh}
with long-range correlated site disorder given by Eq. \eqref{eq:plaw-energies},
all obtained for a single realization of disorder and $L=4096$. (a)
Entanglement spectrum $\zeta_{i}$ obtained from a real- space cut
as a function of the exponent $\alpha$ for $E_{F}=0$. (b) Entanglement
spectrum $\zeta_{i}$ obtained from a cut in momentum space as a function
of the exponent $\alpha$ for $E_{F}=0$. (c) Entanglement spectrum
$\zeta_{i}$ obtained from a cut in momentum space now as a function
of $E_{F}$ for $\alpha=3.0$. (d) Entanglement entropy $S$, divided
by its maximum value $S_{max}$, as a function of the exponent $\alpha$
for cuts both in real and momentum space at $E_{F}=0$. At $E_{F}=0$
extended states are known to exist for $\alpha\gtrsim2.5$,\cite{moura98},
indicated by dashed lines in (a), (b), and (d). For $\alpha=3.0$
extended states exist for Fermi energies in the interval $-0.45\lesssim E_{F}\lesssim0.45$,
indicated by dashed lines in (c). }
\end{figure}


\section{\label{sec:ladder}Short-range correlated ladders}

Encouraged by the sharp distinction between localized and extended
states for $D=1$ systems using the entanglement spectrum and entropy,
as shown in Ref.~\onlinecite{Mondragon-Shem13} and the previous
section, we study now a correlated random ladder model.\cite{sil08}

This tight-binding model consists of two coupled chains of size $L$,
with $N=2L$. Taking advantage of the quasi-$1D$ nature of the problem
we now write the site energies as $2\times2$ matrices
\begin{eqnarray}
\varepsilon_{i} & = & \left(\begin{array}{cc}
\varepsilon_{i,1} & -\gamma_{i}\\
-\gamma_{i} & \varepsilon_{i,2}
\end{array}\right).\label{eq:eps_ladder}
\end{eqnarray}
 where $\gamma_{i}$ is the disordered interchain hopping amplitude
$\gamma_{i}$ and $\varepsilon_{i,1\left(2\right)}$ the on-site energy
at site $i$ and chain $1\left(2\right)$. We consider disorder-free
intrachain hopping amplitude $t=$\unit$t$, where \unit$\,$is
the $2\times2$ identity.

Interestingly, this model shows extended states when the site energies
and the interchain hopping amplitude have a particular correlation:\cite{sil08,moura10}
$\varepsilon_{i,1}=\varepsilon_{i,2}=\gamma_{i}$, with $\varepsilon_{i,1}$
randomly distributed. The emergence these extended states is easily
understood: For this particular choice, the eigenvalues of the $\varepsilon_{i}$
in Eq. \eqref{eq:eps_ladder} are simply $\lambda_{i}=0,\mbox{\,}2\varepsilon_{i,1}$.
Therefore, half of the eigenstates correspond to those of a clean
chain and the other half to those of a chain with uncorrelated on-site
disorder disorder given by $2\varepsilon_{i,1}$.\cite{sil08} Because
of this effective decoupling of the ladder into two chains, we have
the unusual situation where extended states coexist with localized
ones inside the conduction band $-2t\le E\le2t$.

Without loss of generality, we assume that $\varepsilon_{i,1}$ are
uniformly distributed in the interval $\left[-0.5,0.5\right]$ with
$\gamma_{i}=\varepsilon_{i,2}=\varepsilon_{i,1}$. To probe the extent
of the wavefunctions we use the inverse participation ratio, Eq. \eqref{eq:ipr}.
For the present case, it is easy to show that, for the extended states,
$\mbox{IPR}_{\nu}=3/2N$, since these eigenstates are simply the Bloch
states of a clean tight-binding chain. Therefore, for a given realization
of disorder, there are $N/2$ extended eigenstates for which $\mbox{IPR}_{\nu}=3/2N$.

Using the $\mbox{IPR}_{\nu}$ as criterion, we calculate the contributions
to $\rho_{tot}$ for both the extended and localized states, Fig.
\ref{fig:ladder-spectrum}(a). It is then clear that the density of
states of the extended states corresponds to half of the density of
states of a clean tight-binding chain, as expected. For the localized
states, we see that they stretch outside the conduction band $\left(\left|E\right|<2t\right)$,
and that their $\rho_{tot}$ has a similar value as compared to the
density of states for the extended states in the center of the band,
illustrating the advertised coexistence. The novel signatures of this
coexistence were carefully discussed in Ref.~\onlinecite{moura10}.
Here, we complement their results calculating the quantum entanglement.

As in the example of Section~\ref{sec:1d}, we calculate the entanglement
spectrum $\zeta_{i}$ and entropy $S$ for a \emph{single} disorder
realization with partitions both in real and in momentum space. Assuming
that the chains run along the $x$-direction, we perform the cuts
along the points $x=L/2$ and $k_{x}=\pi$, in real and momentum space
respectively. The resulting entanglement entropy and spectrum are
show in Figs. \ref{fig:ladder-spectrum}(b),(c),(d). In distinct difference
to the case discussed in Section~\ref{sec:1d}, here the \emph{only}
signature of extended states is given by a real space cut. This can
be clearly seen in Figs. \ref{fig:ladder-spectrum}(b),(c), where
there is an enhancement of the entanglement for $\left|E_{F}\right|<2t$,
accompanied by its suppression outside this region. Moreover, we notice
that the localized states have minimal effects on this results, since
they give rise to spurious entanglement only for $\left|E_{F}\right|<2t$.
For a momentum-space cut, Figs. \ref{fig:ladder-spectrum}(b),(d),
there is no suppression of the entanglement for all energies for which
$\rho_{tot}>0$.

Naively interpreted this implies that all states for $\left|E_{F}\right|\lesssim2.5t$
are localized. The failure of the momentum-space entanglement to detect
extended states in the current model obviously comes from the fact
that entanglement corresponds to localization and, since localized
states exist for all energies where the extended states are present,
their contribution to $\zeta_{i}$ overcome the lack of entanglement
corresponding to extended states. We notice that more conventional
measures of localization\cite{moura10} correctly capture the existence
of extended states in the current model.

\begin{figure}[t]
\begin{centering}
\includegraphics[scale=0.4]{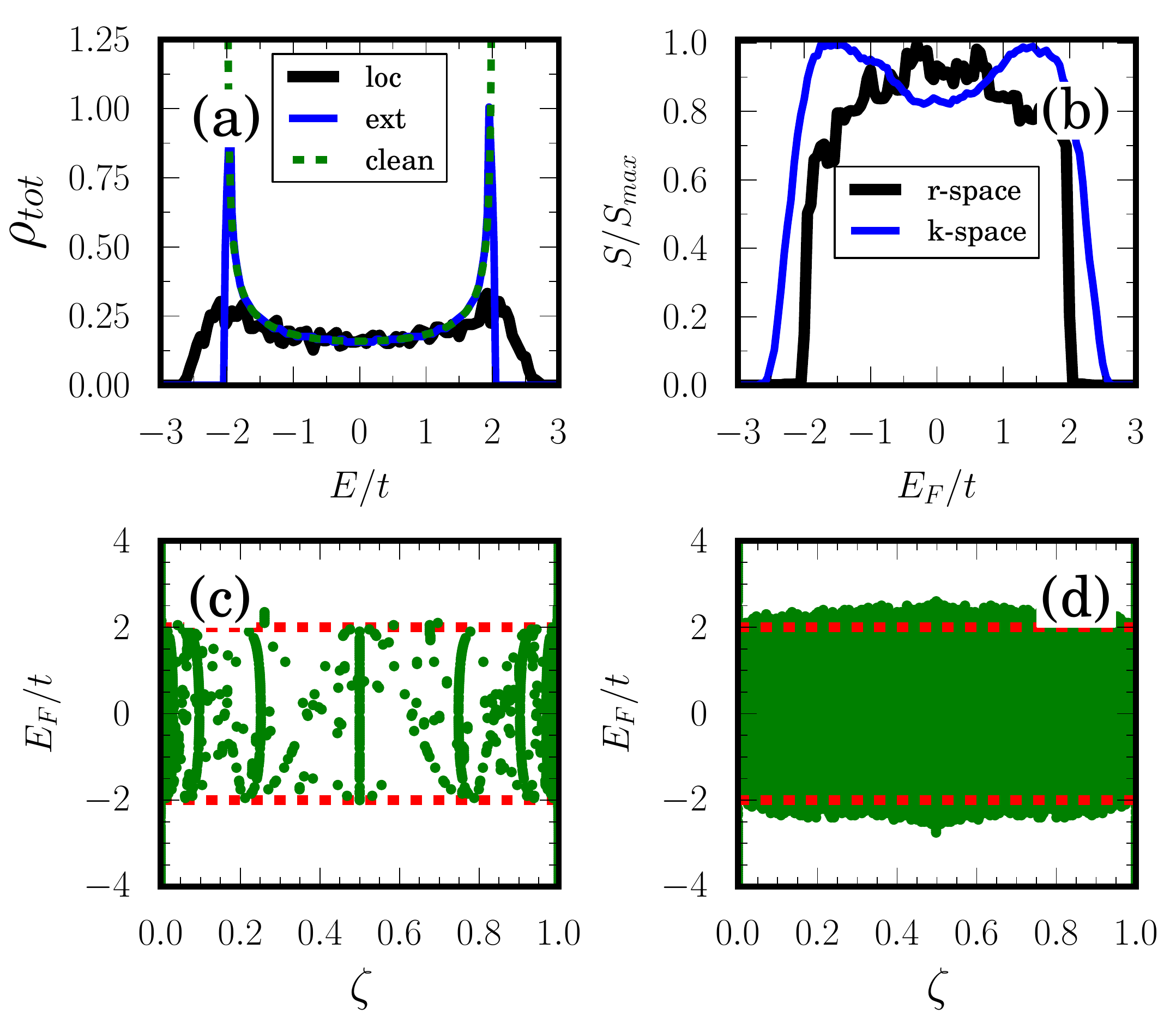}
\par\end{centering}

\caption{\label{fig:ladder-spectrum} (a) Total density of states $\rho_{tot}$
for disordered correlated ladders (Eq. \eqref{eq:eps_ladder}) as
a function of the energy $E$. Using the inverse participation ratio
as criteria to measure the nature of the eigenstates, we plot separately
$\rho_{tot}$ for the localized and extended states and, for comparison,
we plot half of the density of states for a clean tight-binding chain;
(b) Entanglement entropy $S$, for the same model, divided by its
maximum value $S_{max}$, as a function of the Fermi energy $E_{F}$
for cuts both in real and momentum space; (c) Entanglement spectrum
$\zeta_{i}$ obtained from a real space cut as a function of the of
$E_{F}$; (d) Entanglement spectrum $\zeta_{i}$ obtained from a cut
in momentum space as a function of $E_{F}$. The boundary between
localized and extended states is represented by the dashed lines.
The results are obtained for a single realization of disorder for
$N=4096$. }
\end{figure}


\section{\label{sec:2d}Long-range correlated disorder in $D=2$}

Finally, we attempt an extension of the entanglement method to $D=2$.
We consider long-range correlated site energies $\varepsilon_{i,j}$
with spectral density $S(\vec{k})\propto1/|\vec{k}|^{\alpha}$ on
a $L\times L$ square lattice. We follow Ref. ~\onlinecite{santos07}
and define
\begin{eqnarray}
\varepsilon_{i,j} & = & \sum_{n_{i}=1}^{L/2}\sum_{n_{j}=1}^{L/2}\left(i^{2}+j^{2}\right)^{-\alpha/4}\nonumber \\
 & \times & \mbox{cos}\left(\frac{2\pi}{L}n_{i}i+\phi_{i,j}\right)\mbox{cos}\left(\frac{2\pi}{L}n_{j}j+\Omega_{i,j}\right),\label{eq:plaw-energies-2d}
\end{eqnarray}
 where $\phi_{i,j}$ and $\Omega_{i,j}$ are random phases uniformly
distributed in the interval $\left[0,2\pi\right]$ and $k_{x\left(y\right)}=2\pi n_{i\left(j\right)}/L$.
For each value of $\alpha$ we also shift the energies to ensure that
$\left\langle \varepsilon_{i,j}\right\rangle =0$ and normalize them
such that $\sigma_{\varepsilon}=\left\langle \varepsilon_{ij}^{2}\right\rangle =1$.

As in the previous cases, we perform cuts both in real and momentum space and consider a torus geometry, i.e., periodic boundary conditions.\cite{cylinder}
In real space, we perform a partition along the plane $x=L/2$, whereas
in momentum space we cut along the plane $k_{x}=\pi$.

Results are shown in Fig. \ref{fig:plaw-spectrum-2d}. For both partitions
the distinction between localized and extended states is not apparent
in the entanglement spectra and entropy, i.e., unlike for the models discussed
in the previous sections, both methods seem to fail. Of course, we
cannot exclude that the moderate system sizes are simply too small
to probe the nature of the wavefunctions. It is known that careful
finite-size scaling is often necessary to obtain meaningful results
on localization properties.\cite{abrahams79,jia08,tmt,mackinnon81,slevin99,croy12b}
However, we suspect there are further difficulties to the use of the
entanglement spectrum, to be discussed in the next section.

\begin{figure}[t]
\begin{centering}
\includegraphics[scale=0.4]{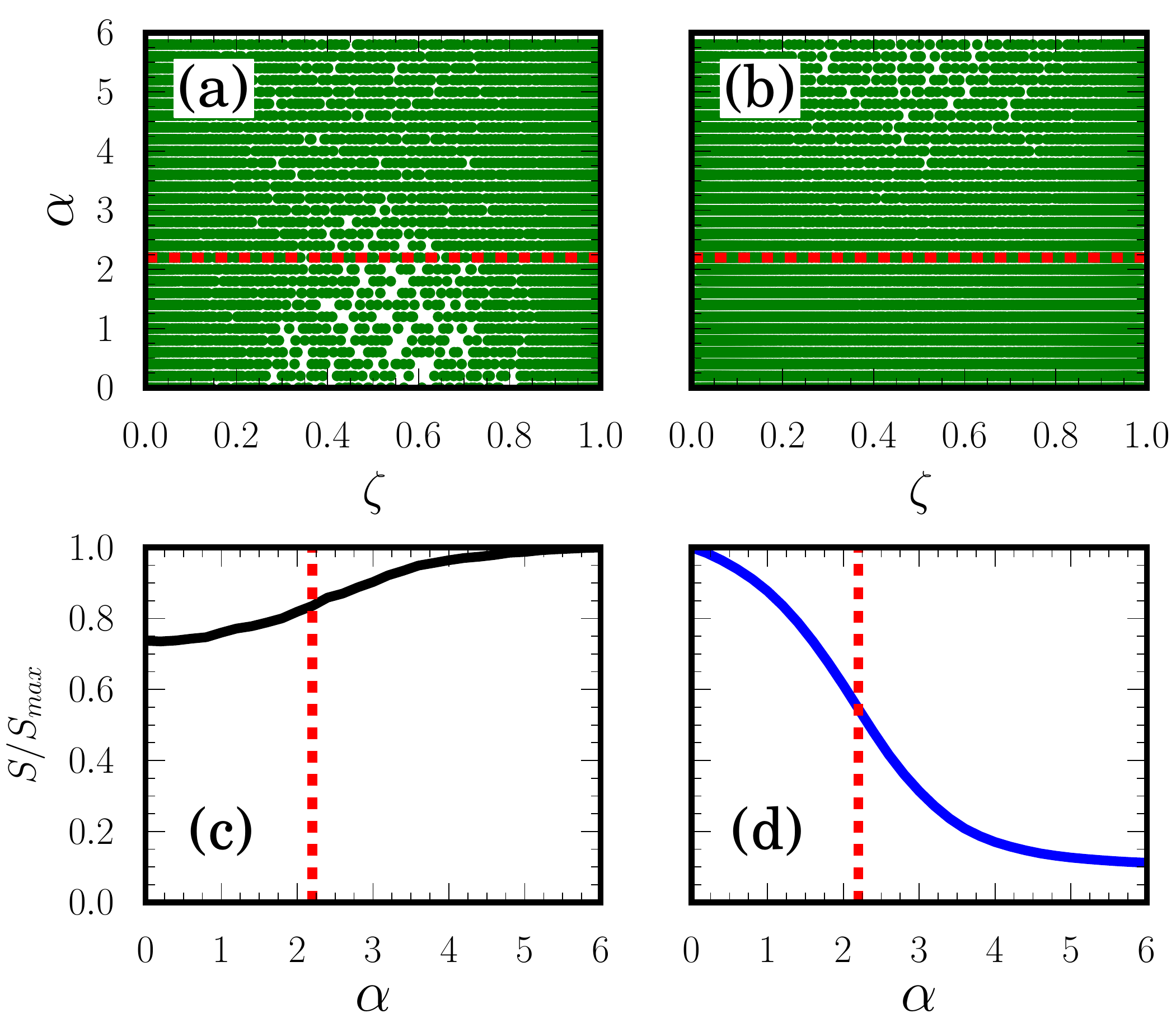}
\par\end{centering}

\caption{\label{fig:plaw-spectrum-2d} Entanglement spectrum $\zeta_{i}$ and
entropy $S$ for the disordered tight-binding Hamiltonian Eq. \eqref{eq:tbh}
with long-range correlated site disorder, as given by Eq. \eqref{eq:plaw-energies-2d},
in $D=2$. (a) Entanglement spectrum $\zeta_{i}$ obtained from a
real space cut as a function of the exponent $\alpha$ for $E_{F}=-2.5t$;
(b) Entanglement spectrum $\zeta_{i}$ obtained from a cut in momentum space
as a function of the exponent $\alpha$ for $E_{F}=-2.5t$. Entanglement
entropy $S$, divided by its maximum value $S_{max}$, as a function
of the exponent $\alpha$ for $E_{F}=-2.5t$ considering cuts both
in real (c) and momentum space (d). Extend states exist for $\alpha\gtrsim2.2$,\cite{santos07}
and the boundary between localized and extended states is represented
by the dashed lines. The results are obtained for a single realization
of disorder in a square lattice of linear size $L=96$.}
\end{figure}


\section{\label{sec:discussion}Analysis and discussion}

Given that we have encountered only limited success of the momentum-space
entanglement method proposed in Ref.~\onlinecite{Mondragon-Shem13},
a more in-depth discussion of its working principles is required which
we attempt here.

\subsection{Correlation matrix}

As mentioned in Sec. \ref{sec:model-entang}, the correlation matrix
$C_{ij}$ has contributions from {\em all} states below the Fermi
level $E_{F}$. Naively, one would expect that increasing $E_{F}$
therefore simply adds new eigenvalues to $C_{ij}$. In the presence
of a mobility edge at $E_{c}$, with exclusively localized states
below $E_{c}$, one might therefore expect that these localized states
always contribute eigenvalues $0<\zeta_{i}<1$ in the momentum-space
partition even for $E_{F}>E_{c}$. However, this is clearly not the
case, as shown in Fig. \ref{fig:plaw-spectrum-1d}(c) where such eigenvalues
do not occur in an energy window between the mobility edges.

To settle this apparent contradiction, we have to remember the fact
that we are dealing with long-range correlated disorder. While a generic
(short-range) disorder spreads single-particle states over all of
momentum space, long-range disorder spreads the states only over a
much narrower $k$ interval. Suppose now that there are localized
states in some energy interval $E_{1}<E<E_{2}$. In the clean case,
these energies can be assigned to some momenta $k_{1}<\left|k\right|<k_{2}$.
If now $E_{F}$ is sufficiently far above $E_{2}$ then all momenta
near $k_{1}$, $k_{2}$ will be totally filled (despite the presence
of localized states), because momentum smearing is moderate. Therefore,
the non-vanishing $\zeta_{i}$ will only be $1$, regardless of the fact
that there are localized states below $E_{F}$. Consequently, the
energies for which momentum-space entanglement, as calculated from
$C_{ij}$, is suppressed do not coincide with the true mobility edges,
such that $\zeta_{i}$ cannot, by construction, sharply determine
the position of mobility edges. Notice that this discussion does not
apply to the model considered in Ref.~\onlinecite{Mondragon-Shem13}
because this only shows extended states at a particular resonant energy
and thus has no mobility edges.\cite{dunlap90}

We conclude that the entanglement spectrum constructed from $C_{ij}$,
which has contributions from all states below $E_{F}$, renders an
energy-resolved detection of localization problematic; this becomes
more pressing for short-range correlated disorder and is therefore
of relevance for generalizations to $D\ge2$. The problem may be resolved
by a different construction of an appropriate entanglement spectrum.
We recall, however, that the Fermi-sea construction of $C_{ij}$ is
vital for momentum-space partitions in order to remove the trivial
entanglement between $k$ and $-k$ of extended states.

\subsection{Spurious entanglement}

We believe that the apparent failure of the entanglement measures
in the $D=2$ case, Fig.~\ref{fig:plaw-spectrum-2d}, is primarily
related to spurious entanglement. It is well known that the entanglement
entropy has the general property to follow an ``area law'',\cite{srednicki93}
i.e., to scale according to $L^{D-1}$ in a system of linear dimension
$L$, simply because the cut used to partition the system scales as
$L^{D-1}$. In other words, for $D\ge2$ there is sizeable entanglement
across the boundary between $\mathcal{A}$ and $\mathcal{B}$ whose
nature is trivial but which contributes eigenvalues $0<\zeta_{i}<1$
of $C_{ij}$ which spoil the naive analysis of the entanglement spectrum.

Hence, the challenge lies in separating this spurious (or short-range)
entanglement from the long-range entanglement which contains the physics
of localization vs. delocalization we are interested in. Such a separation
very likely requires a careful and detailed analysis the system-size
dependence of entanglement properties, as has been done in recent
numerical work on spin systems.\cite{furukawa07,isakov11}

\section{\label{sec:summary}Summary}

Using the recently proposed\cite{Mondragon-Shem13} concept of momentum-space
entanglement, we studied Anderson localization and the associated
metal--insulator transition in low-dimensional tight-binding models
with spatially correlated disorder.

Similar to the results for a $D=1$ random-dimer model in Ref.~\onlinecite{Mondragon-Shem13},
we find for a $D=1$ model with long-range disorder correlations that
the momentum-space entanglement calculated via the correlation matrix
appears to provide an efficient differentiation between localized
and extended states, even for a single realization of disorder. In
contrast, for a tight-binding ladder with special short-range disorder
correlations, which features coexisting extended and localized states
over a range of energies, momentum-space entanglement is not easily
able to detect the presence of extended states. Finally, an extension
to $D=2$ which we applied to a model with long-range correlated disorder
appears to fail entirely: We found no obvious signatures of the Anderson transition
both for a momentum-space and a real-space partition.

A general discussion of the method indicates two reasons for failure,
one related to the many-particle character of the correlation matrix
and the other one related to spurious entanglement. We conclude that
more elaborate extensions of the method proposed in Ref.~\onlinecite{Mondragon-Shem13}
are required to access localization in $D\ge2$. Such extensions might
require an energy-selective definition of the correlation matrix and
a more quantitative analysis of entanglement spectrum and entropy,
combined with finite-size scaling, in order to separate spurious
(boundary) entanglement from long-range entanglement. It remains to
be seen whether such an extended method still provides advantages
over more conventional measures of localization.

\acknowledgements

This research was supported by the DFG through FOR 960 and GRK 1621.
ECA was also partially supported by FAPESP. We kindly acknowledge
ZIH at TU Dresden for the allocation of computing resources.



\begin{thebibliography}{39}

\expandafter\ifx\csname natexlab\endcsname\relax\def\natexlab#1{#1}\fi
\expandafter\ifx\csname bibnamefont\endcsname\relax
  \def\bibnamefont#1{#1}\fi
\expandafter\ifx\csname bibfnamefont\endcsname\relax
  \def\bibfnamefont#1{#1}\fi
\expandafter\ifx\csname citenamefont\endcsname\relax
  \def\citenamefont#1{#1}\fi
\expandafter\ifx\csname url\endcsname\relax
  \def\url#1{\texttt{#1}}\fi
\expandafter\ifx\csname urlprefix\endcsname\relax\def\urlprefix{URL }\fi
\providecommand{\bibinfo}[2]{#2}
\providecommand{\eprint}[2][]{\url{#2}}

\bibitem[{\citenamefont{Anderson}(1958)}]{anderson58}
\bibinfo{author}{\bibfnamefont{P.~W.} \bibnamefont{Anderson}},
  \bibinfo{journal}{Phys. Rev.} \textbf{\bibinfo{volume}{109}},
  \bibinfo{pages}{1492} (\bibinfo{year}{1958}).

\bibitem[{\citenamefont{Dobrosavljevi{\'c}
  et~al.}(2012)\citenamefont{Dobrosavljevi{\'c}, Trivedi, and {Valles
  Jr.}}}]{book-vlad}
\bibinfo{editor}{\bibfnamefont{V.}~\bibnamefont{Dobrosavljevi{\'c}}},
  \bibinfo{editor}{\bibfnamefont{N.}~\bibnamefont{Trivedi}}, \bibnamefont{and}
  \bibinfo{editor}{\bibfnamefont{J.~M.} \bibnamefont{{Valles Jr.}}}, eds.,
  \emph{\bibinfo{title}{{Conductor Insulator Quantum Phase Transitions}}}
  (\bibinfo{publisher}{Oxford University Press}, \bibinfo{address}{Oxford},
  \bibinfo{year}{2012}).

\bibitem[{\citenamefont{Abrahams et~al.}(1979)\citenamefont{Abrahams, Anderson,
  Licciardello, and Ramakrishnan}}]{abrahams79}
\bibinfo{author}{\bibfnamefont{E.}~\bibnamefont{Abrahams}},
  \bibinfo{author}{\bibfnamefont{P.~W.} \bibnamefont{Anderson}},
  \bibinfo{author}{\bibfnamefont{D.~C.} \bibnamefont{Licciardello}},
  \bibnamefont{and} \bibinfo{author}{\bibfnamefont{T.~V.}
  \bibnamefont{Ramakrishnan}}, \bibinfo{journal}{Phys. Rev. Lett.}
  \textbf{\bibinfo{volume}{42}}, \bibinfo{pages}{673} (\bibinfo{year}{1979}).

\bibitem[{\citenamefont{Dunlap et~al.}(1990)\citenamefont{Dunlap, Wu, and
  Phillips}}]{dunlap90}
\bibinfo{author}{\bibfnamefont{D.~H.} \bibnamefont{Dunlap}},
  \bibinfo{author}{\bibfnamefont{H.-L.} \bibnamefont{Wu}}, \bibnamefont{and}
  \bibinfo{author}{\bibfnamefont{P.~W.} \bibnamefont{Phillips}},
  \bibinfo{journal}{Phys. Rev. Lett.} \textbf{\bibinfo{volume}{65}},
  \bibinfo{pages}{88} (\bibinfo{year}{1990}).

\bibitem[{\citenamefont{de~Moura and Lyra}(1998)}]{moura98}
\bibinfo{author}{\bibfnamefont{F.~A. B.~F.} \bibnamefont{de~Moura}}
  \bibnamefont{and} \bibinfo{author}{\bibfnamefont{M.~L.} \bibnamefont{Lyra}},
  \bibinfo{journal}{Phys. Rev. Lett.} \textbf{\bibinfo{volume}{81}},
  \bibinfo{pages}{3735} (\bibinfo{year}{1998}).

\bibitem[{\citenamefont{Izrailev and Krokhin}(1999)}]{izrailev99}
\bibinfo{author}{\bibfnamefont{F.~M.} \bibnamefont{Izrailev}} \bibnamefont{and}
  \bibinfo{author}{\bibfnamefont{A.~A.} \bibnamefont{Krokhin}},
  \bibinfo{journal}{Phys. Rev. Lett.} \textbf{\bibinfo{volume}{82}},
  \bibinfo{pages}{4062} (\bibinfo{year}{1999}).

\bibitem[{\citenamefont{Garc{\'i}a-Garc{\'i}a and
  Cuevas}(2009)}]{garcia-garcia09}
\bibinfo{author}{\bibfnamefont{A.~M.} \bibnamefont{Garc{\'i}a-Garc{\'i}a}}
  \bibnamefont{and} \bibinfo{author}{\bibfnamefont{E.}~\bibnamefont{Cuevas}},
  \bibinfo{journal}{Phys. Rev. B} \textbf{\bibinfo{volume}{79}},
  \bibinfo{pages}{073104} (\bibinfo{year}{2009}).

\bibitem[{\citenamefont{dos Santos et~al.}(2007)\citenamefont{dos Santos,
  de~Moura, Lyra, and Coutinho-Filho}}]{santos07}
\bibinfo{author}{\bibfnamefont{I.~F.} \bibnamefont{dos Santos}},
  \bibinfo{author}{\bibfnamefont{F.~A. B.~F.} \bibnamefont{de~Moura}},
  \bibinfo{author}{\bibfnamefont{M.~L.} \bibnamefont{Lyra}}, \bibnamefont{and}
  \bibinfo{author}{\bibfnamefont{M.~D.} \bibnamefont{Coutinho-Filho}},
  \bibinfo{journal}{J. Phys.: Condens. Matter} \textbf{\bibinfo{volume}{19}},
  \bibinfo{pages}{476213} (\bibinfo{year}{2007}).

\bibitem[{\citenamefont{Sil et~al.}(2008)\citenamefont{Sil, Maiti, and
  Chakrabarti}}]{sil08}
\bibinfo{author}{\bibfnamefont{S.}~\bibnamefont{Sil}},
  \bibinfo{author}{\bibfnamefont{S.~K.} \bibnamefont{Maiti}}, \bibnamefont{and}
  \bibinfo{author}{\bibfnamefont{A.}~\bibnamefont{Chakrabarti}},
  \bibinfo{journal}{Phys. Rev. B} \textbf{\bibinfo{volume}{78}},
  \bibinfo{pages}{113103} (\bibinfo{year}{2008}).

\bibitem[{\citenamefont{Croy et~al.}(2011)\citenamefont{Croy, Cain, and
  Schreiber}}]{croy11}
\bibinfo{author}{\bibfnamefont{A.}~\bibnamefont{Croy}},
  \bibinfo{author}{\bibfnamefont{P.}~\bibnamefont{Cain}}, \bibnamefont{and}
  \bibinfo{author}{\bibfnamefont{M.}~\bibnamefont{Schreiber}},
  \bibinfo{journal}{Eur. Phys. J. B} \textbf{\bibinfo{volume}{82}},
  \bibinfo{pages}{107} (\bibinfo{year}{2011}).

\bibitem[{\citenamefont{Herbut}(2001)}]{herbut01}
\bibinfo{author}{\bibfnamefont{I.~F.} \bibnamefont{Herbut}},
  \bibinfo{journal}{Phys. Rev. B} \textbf{\bibinfo{volume}{63}},
  \bibinfo{pages}{113102} (\bibinfo{year}{2001}).

\bibitem[{\citenamefont{Andrade et~al.}(2010)\citenamefont{Andrade, Miranda,
  and {Dobrosavljevi\ifmmode \acute{c}\else {\'c}\fi{}}}}]{andrade10}
\bibinfo{author}{\bibfnamefont{E.~C.} \bibnamefont{Andrade}},
  \bibinfo{author}{\bibfnamefont{E.}~\bibnamefont{Miranda}}, \bibnamefont{and}
  \bibinfo{author}{\bibfnamefont{V.}~\bibnamefont{{Dobrosavljevi\ifmmode
  \acute{c}\else {\'c}\fi{}}}}, \bibinfo{journal}{Phys. Rev. Lett.}
  \textbf{\bibinfo{volume}{104}}, \bibinfo{pages}{236401}
  (\bibinfo{year}{2010}).

\bibitem[{\citenamefont{Dobrosavljevi{\'c}
  et~al.}(2003)\citenamefont{Dobrosavljevi{\'c}, Pastor, and
  Nikoli{\'c}}}]{tmt}
\bibinfo{author}{\bibfnamefont{V.}~\bibnamefont{Dobrosavljevi{\'c}}},
  \bibinfo{author}{\bibfnamefont{A.~A.} \bibnamefont{Pastor}},
  \bibnamefont{and} \bibinfo{author}{\bibfnamefont{B.~K.}
  \bibnamefont{Nikoli{\'c}}}, \bibinfo{journal}{Europhys. Lett.}
  \textbf{\bibinfo{volume}{62}}, \bibinfo{pages}{76} (\bibinfo{year}{2003}).

\bibitem[{\citenamefont{Wegner}(1980)}]{wegner80}
\bibinfo{author}{\bibfnamefont{F.}~\bibnamefont{Wegner}}, \bibinfo{journal}{Z.
  Phys. B} \textbf{\bibinfo{volume}{36}}, \bibinfo{pages}{209}
  (\bibinfo{year}{1980}).

\bibitem[{\citenamefont{Johri and Bhatt}(2012)}]{johri12}
\bibinfo{author}{\bibfnamefont{S.}~\bibnamefont{Johri}} \bibnamefont{and}
  \bibinfo{author}{\bibfnamefont{R.~N.} \bibnamefont{Bhatt}},
  \bibinfo{journal}{Phys. Rev. Lett.} \textbf{\bibinfo{volume}{109}},
  \bibinfo{pages}{076402} (\bibinfo{year}{2012}).

\bibitem[{\citenamefont{Farchioni et~al.}(1992)\citenamefont{Farchioni, Grosso,
  and {Pastori Parravicini}}}]{farchioni92}
\bibinfo{author}{\bibfnamefont{R.}~\bibnamefont{Farchioni}},
  \bibinfo{author}{\bibfnamefont{G.}~\bibnamefont{Grosso}}, \bibnamefont{and}
  \bibinfo{author}{\bibfnamefont{G.}~\bibnamefont{{Pastori Parravicini}}},
  \bibinfo{journal}{Phys. Rev. B} \textbf{\bibinfo{volume}{45}},
  \bibinfo{pages}{6383} (\bibinfo{year}{1992}).

\bibitem[{\citenamefont{MacKinnon and Kramer}(1981)}]{mackinnon81}
\bibinfo{author}{\bibfnamefont{A.}~\bibnamefont{MacKinnon}} \bibnamefont{and}
  \bibinfo{author}{\bibfnamefont{B.}~\bibnamefont{Kramer}},
  \bibinfo{journal}{Phys. Rev. Lett.} \textbf{\bibinfo{volume}{47}},
  \bibinfo{pages}{1546} (\bibinfo{year}{1981}).

\bibitem[{\citenamefont{Anderson et~al.}(1980)\citenamefont{Anderson, Thouless,
  Abrahams, and Fisher}}]{anderson80}
\bibinfo{author}{\bibfnamefont{P.~W.} \bibnamefont{Anderson}},
  \bibinfo{author}{\bibfnamefont{D.~J.} \bibnamefont{Thouless}},
  \bibinfo{author}{\bibfnamefont{E.}~\bibnamefont{Abrahams}}, \bibnamefont{and}
  \bibinfo{author}{\bibfnamefont{D.~S.} \bibnamefont{Fisher}},
  \bibinfo{journal}{Phys. Rev. B} \textbf{\bibinfo{volume}{22}},
  \bibinfo{pages}{3519} (\bibinfo{year}{1980}).

\bibitem[{\citenamefont{Marko\v{s}}(2006)}]{markos06}
\bibinfo{author}{\bibfnamefont{P.}~\bibnamefont{Marko\v{s}}},
  \bibinfo{journal}{Acta Phys. Slovaca} \textbf{\bibinfo{volume}{51}},
  \bibinfo{pages}{581} (\bibinfo{year}{2006}).

\bibitem[{\citenamefont{Jia et~al.}(2008)\citenamefont{Jia, Subramaniam,
  Gruzberg, and Chakravarty}}]{jia08}
\bibinfo{author}{\bibfnamefont{X.}~\bibnamefont{Jia}},
  \bibinfo{author}{\bibfnamefont{A.~R.} \bibnamefont{Subramaniam}},
  \bibinfo{author}{\bibfnamefont{I.~A.} \bibnamefont{Gruzberg}},
  \bibnamefont{and}
  \bibinfo{author}{\bibfnamefont{S.}~\bibnamefont{Chakravarty}},
  \bibinfo{journal}{Phys. Rev. B} \textbf{\bibinfo{volume}{77}},
  \bibinfo{pages}{014208} (\bibinfo{year}{2008}).

\bibitem[{\citenamefont{Chen et~al.}(2012)\citenamefont{Chen, Hsu, Hughes, and
  Fradkin}}]{chen12}
\bibinfo{author}{\bibfnamefont{X.}~\bibnamefont{Chen}},
  \bibinfo{author}{\bibfnamefont{B.}~\bibnamefont{Hsu}},
  \bibinfo{author}{\bibfnamefont{T.~L.} \bibnamefont{Hughes}},
  \bibnamefont{and} \bibinfo{author}{\bibfnamefont{E.}~\bibnamefont{Fradkin}},
  \bibinfo{journal}{Phys. Rev. B} \textbf{\bibinfo{volume}{86}},
  \bibinfo{pages}{134201} (\bibinfo{year}{2012}).

\bibitem[{\citenamefont{Mondragon-Shem
  et~al.}(2013)\citenamefont{Mondragon-Shem, Khan, and
  Hughes}}]{Mondragon-Shem13}
\bibinfo{author}{\bibfnamefont{I.}~\bibnamefont{Mondragon-Shem}},
  \bibinfo{author}{\bibfnamefont{M.}~\bibnamefont{Khan}}, \bibnamefont{and}
  \bibinfo{author}{\bibfnamefont{T.~L.} \bibnamefont{Hughes}},
  \bibinfo{journal}{Phys. Rev. Lett.} \textbf{\bibinfo{volume}{110}},
  \bibinfo{pages}{046806} (\bibinfo{year}{2013}).

\bibitem[{\citenamefont{Thomale et~al.}(2010)\citenamefont{Thomale, Arovas, and
  Bernevig}}]{thomale10}
\bibinfo{author}{\bibfnamefont{R.}~\bibnamefont{Thomale}},
  \bibinfo{author}{\bibfnamefont{D.~P.} \bibnamefont{Arovas}},
  \bibnamefont{and} \bibinfo{author}{\bibfnamefont{B.~A.}
  \bibnamefont{Bernevig}}, \bibinfo{journal}{Phys. Rev. Lett.}
  \textbf{\bibinfo{volume}{105}}, \bibinfo{pages}{116805}
  (\bibinfo{year}{2010}).

\bibitem[{\citenamefont{Lundgren et~al.}(2012)\citenamefont{Lundgren, Chua, and
  Fiete}}]{lundgren12}
\bibinfo{author}{\bibfnamefont{R.}~\bibnamefont{Lundgren}},
  \bibinfo{author}{\bibfnamefont{V.}~\bibnamefont{Chua}}, \bibnamefont{and}
  \bibinfo{author}{\bibfnamefont{G.~A.} \bibnamefont{Fiete}},
  \bibinfo{journal}{Phys. Rev. B} \textbf{\bibinfo{volume}{86}},
  \bibinfo{pages}{224422} (\bibinfo{year}{2012}).

\bibitem[{\citenamefont{Peschel}(2003)}]{peschel03}
\bibinfo{author}{\bibfnamefont{I.}~\bibnamefont{Peschel}}, \bibinfo{journal}{J.
  Phys. A: Math. Gen.} \textbf{\bibinfo{volume}{36}}, \bibinfo{pages}{L205}
  (\bibinfo{year}{2003}).

\bibitem[{\citenamefont{Peschel and Eisler}(2009)}]{peschel09}
\bibinfo{author}{\bibfnamefont{I.}~\bibnamefont{Peschel}} \bibnamefont{and}
  \bibinfo{author}{\bibfnamefont{V.}~\bibnamefont{Eisler}},
  \bibinfo{journal}{J. Phys. A: Math. Theor.} \textbf{\bibinfo{volume}{42}},
  \bibinfo{pages}{504003} (\bibinfo{year}{2009}).

\bibitem[{\citenamefont{Pouranvari and Yang}(2014)}]{pouranvari14}
\bibinfo{author}{\bibfnamefont{M.}~\bibnamefont{Pouranvari}} \bibnamefont{and}
  \bibinfo{author}{\bibfnamefont{K.}~\bibnamefont{Yang}},
  \bibinfo{journal}{Phys. Rev. B} \textbf{\bibinfo{volume}{89}},
  \bibinfo{pages}{115104} (\bibinfo{year}{2014}).

\bibitem[{\citenamefont{Prodan et~al.}(2010)\citenamefont{Prodan, Hughes, and
  Bernevig}}]{prodan10}
\bibinfo{author}{\bibfnamefont{E.}~\bibnamefont{Prodan}},
  \bibinfo{author}{\bibfnamefont{T.~L.} \bibnamefont{Hughes}},
  \bibnamefont{and} \bibinfo{author}{\bibfnamefont{B.~A.}
  \bibnamefont{Bernevig}}, \bibinfo{journal}{Phys. Rev. Lett.}
  \textbf{\bibinfo{volume}{105}}, \bibinfo{pages}{115501}
  (\bibinfo{year}{2010}).

\bibitem[{\citenamefont{Mondragon-Shem and Hughes}(2014)}]{Mondragon-Shem14}
\bibinfo{author}{\bibfnamefont{I.}~\bibnamefont{Mondragon-Shem}}
  \bibnamefont{and} \bibinfo{author}{\bibfnamefont{T.~L.}
  \bibnamefont{Hughes}}, \bibinfo{journal}{arXiv:1403.6129v1 [cond-mat.dis-nn]}
   (\bibinfo{year}{2014}).

\bibitem[{\citenamefont{Osborne and Provenzale}(1989)}]{osborne89}
\bibinfo{author}{\bibfnamefont{A.}~\bibnamefont{Osborne}} \bibnamefont{and}
  \bibinfo{author}{\bibfnamefont{A.}~\bibnamefont{Provenzale}},
  \bibinfo{journal}{Phys. D} \textbf{\bibinfo{volume}{35}},
  \bibinfo{pages}{357} (\bibinfo{year}{1989}).

\bibitem[{\citenamefont{Petersen and Sandler}(2013)}]{petersen13}
\bibinfo{author}{\bibfnamefont{G.~M.} \bibnamefont{Petersen}} \bibnamefont{and}
  \bibinfo{author}{\bibfnamefont{N.}~\bibnamefont{Sandler}},
  \bibinfo{journal}{Phys. Rev. B} \textbf{\bibinfo{volume}{87}},
  \bibinfo{pages}{195443} (\bibinfo{year}{2013}).

\bibitem[{\citenamefont{Nishino et~al.}(2009)\citenamefont{Nishino, Yakubo, and
  Shima}}]{nishino09}
\bibinfo{author}{\bibfnamefont{S.}~\bibnamefont{Nishino}},
  \bibinfo{author}{\bibfnamefont{K.}~\bibnamefont{Yakubo}}, \bibnamefont{and}
  \bibinfo{author}{\bibfnamefont{H.}~\bibnamefont{Shima}},
  \bibinfo{journal}{Phys. Rev. B} \textbf{\bibinfo{volume}{79}},
  \bibinfo{pages}{033105} (\bibinfo{year}{2009}).

\bibitem[{\citenamefont{de~Moura et~al.}(2010)\citenamefont{de~Moura, Caetano,
  and Lyra}}]{moura10}
\bibinfo{author}{\bibfnamefont{F.~A. B.~F.} \bibnamefont{de~Moura}},
  \bibinfo{author}{\bibfnamefont{R.~A.} \bibnamefont{Caetano}},
  \bibnamefont{and} \bibinfo{author}{\bibfnamefont{M.~L.} \bibnamefont{Lyra}},
  \bibinfo{journal}{Phys. Rev. B} \textbf{\bibinfo{volume}{81}},
  \bibinfo{pages}{125104} (\bibinfo{year}{2010}).

\bibitem[{cyl()}]{cylinder}
\bibinfo{note}{We also considered a cylinder geometry with similar results.}

\bibitem[{\citenamefont{Slevin and Ohtsuki}(1999)}]{slevin99}
\bibinfo{author}{\bibfnamefont{K.}~\bibnamefont{Slevin}} \bibnamefont{and}
  \bibinfo{author}{\bibfnamefont{T.}~\bibnamefont{Ohtsuki}},
  \bibinfo{journal}{Phys. Rev. Lett.} \textbf{\bibinfo{volume}{82}},
  \bibinfo{pages}{382} (\bibinfo{year}{1999}).

\bibitem[{\citenamefont{Croy and Schreiber}(2012)}]{croy12b}
\bibinfo{author}{\bibfnamefont{A.}~\bibnamefont{Croy}} \bibnamefont{and}
  \bibinfo{author}{\bibfnamefont{M.}~\bibnamefont{Schreiber}},
  \bibinfo{journal}{Phys. Rev. B} \textbf{\bibinfo{volume}{85}},
  \bibinfo{pages}{205147} (\bibinfo{year}{2012}).

\bibitem[{\citenamefont{Srednicki}(1993)}]{srednicki93}
\bibinfo{author}{\bibfnamefont{M.}~\bibnamefont{Srednicki}},
  \bibinfo{journal}{Phys. Rev. Lett.} \textbf{\bibinfo{volume}{71}},
  \bibinfo{pages}{666} (\bibinfo{year}{1993}).

\bibitem[{\citenamefont{Furukawa and Misguich}(2007)}]{furukawa07}
\bibinfo{author}{\bibfnamefont{S.}~\bibnamefont{Furukawa}} \bibnamefont{and}
  \bibinfo{author}{\bibfnamefont{G.}~\bibnamefont{Misguich}},
  \bibinfo{journal}{Phys. Rev. B} \textbf{\bibinfo{volume}{75}},
  \bibinfo{pages}{214407} (\bibinfo{year}{2007}).

\bibitem[{\citenamefont{Isakov et~al.}(2011)\citenamefont{Isakov, Hastings, and
  Melko}}]{isakov11}
\bibinfo{author}{\bibfnamefont{S.~V.} \bibnamefont{Isakov}},
  \bibinfo{author}{\bibfnamefont{M.~B.} \bibnamefont{Hastings}},
  \bibnamefont{and} \bibinfo{author}{\bibfnamefont{R.~G.} \bibnamefont{Melko}},
  \bibinfo{journal}{Nature Phys.} \textbf{\bibinfo{volume}{7}},
  \bibinfo{pages}{772} (\bibinfo{year}{2011}).

\end{thebibliography}
\end{document}